# A comparison of two suffix tree-based document clustering algorithms

Muhammad Rafi, Mehdi Maujood ,Murtaza Munawar Fazal, Syed Muhammad Ali
{ muhammad.rafi | k060069 | k060195 | k060062} @nu.edu.pk
National University of Computer & Emerging Sciences
NU-FAST, Karachi, Pakistan

*Abstract*— Document clustering as an unsupervised approach extensively used to navigate, filter, summarize and manage large collection of document repositories like the World Wide Web (WWW). Recently, focuses in this domain shifted from traditional vector based document similarity for clustering to suffix tree based document similarity, as it offers more semantic representation of the text present in the document. In this paper, we compare and contrast two recently introduced approaches to document clustering based on suffix tree data model. The first is an Efficient Phrase based document clustering, which extracts phrases from documents to form compact document representation and uses a similarity measure based on common suffix tree to cluster the documents. The second approach is a frequent word/word meaning sequence based document clustering, it similarly extracts the common word sequence from the document and uses the common sequence/ common word meaning sequence to perform the compact representation, and finally, it uses document clustering approach to cluster the compact documents. These algorithms are using agglomerative hierarchical document clustering to perform the actual clustering step, the difference in these approaches are mainly based on extraction of phrases, model representation as a compact document, and the similarity measures used for clustering. This paper investigates the computational aspect of the two algorithms, and the quality of results they produced.

I. INTRODUCTION

Data clustering is an unsupervised data mining approach that groups a large collection of objects (data points, records, entities, documents etc.) in to more meaning -full smaller sub-groups. This process collects the largely similar objects in some sense into a distinct group, while the largely dissimilar objects in the same sense into different groups (cluster). Document clustering, is a specialized data clustering problem, where the objects are in the form of documents. The objective of the clustering process is to group the documents which are similar in some sense like: type of document, contents of document, etc into a single group (cluster).
The problem of document clustering can be stated as follow:

Given a document collection $D=\{d_1,d_2,d_3,....d_N\}$ which contains N documents, we need to sub-group the documents based on the semantic of the text contents present in a Document , assuming we require K such sub-groups, the clustering process generates $C=\{c_1,c_2,...c_k\}$ clusters, with each $c_i$ being non empty.

Document clustering is still a developing field which is undergoing evolution. It started off on the popular vector based approach where documents were treated as a bag of words and clustering criteria was the presence of common words in the documents. Several modifications were applied on this method to improve this method as the result set would only provide us information on what words were present in a group of documents, not the actual content or context of the documents. There was a need of more intuitive ways of clustering that would provide us sound knowledge of the content present inside the documents.

Semantic document clustering provides a means of clustering documents on the basis of the actual content inside the document. It exploits the semantics present inside the documents (phrases, frequent words, sentence structures etc) as a criterion function to cluster documents. Semantic clustering usually employs the "suffix tree" data structure as it best preserves the order of the words in the sentences present inside the document.

We compare two approaches that utilize the Suffix Tree Data Model; the New Suffix Tree Clustering (NSTC) algorithm proposed by Chim and Deng [11], and Clustering Based on Frequent Word/Word Meaning Sequences (CFWS/WMS) proposed by Yanjun Li et al. [12]. These two experiments reported encouraging results in corresponding authors' experiments, but have not been compared against each other. In the next section, we discuss the related work of this study, and then we discuss the experimental setup, data set and the results of this study. Finally, in the last section we discuss the conclusion and future work.

## II. RELATED WORK

Clustering as an unsupervised machine learning method, is an effective data mining technique that has been comprehensively studied and extensively applied to a variety of application areas. A detail on different data clustering approaches can be found in [1]. Document clustering is a special data clustering technique that clusters a document collection into meaningful sub-collections. The documents in each sub-collection are highly related (in some sense) to each other and vastly different (in the same sense) to the documents placed in other sub-collections. The ever-increasing number of documents in public or corporate collections like: World Wide Web (www) and document repositories at corporate intranet; encourage researchers to finds ways to handle this information overload. Clustering is an effective method for search computing [2]. It offers the possibilities like: grouping similar results [3], comprehend the links between the results [4] and creating the succinct representation and display of search results.

Document clustering problem received a lot of attention from the research community recently. There are two major categories of clustering algorithms that are applied to document clustering (i) Hierarchical vs. Flat and (ii) Partition vs. Overlapping. Agglomerative hierarchical clustering [1] (AHC) initially treats each document as a cluster, and compute a similarity measure for every pair of document, a variety of similarity measures [5] have been utilized for this algorithm. The calculations for all pairs of document similarity demand a lot of computation. The AHC-algorithm then merges the two closest pair iteratively to produce the desirable number of clusters. This is a bottom-up approach to AHC, a top-down approach is also feasible and various modified algorithms are also suggested by various researchers. Unweighted Pair group method with Arithmetic Mean (UPGMA), a derivative of AHC is reported as the best algorithm in this category. The second category of document clustering algorithms is partitioned based algorithms, [3] [6] [7] which create a one level partitioning of the document collection. The k-means algorithm initially selects k-documents as seed, these documents are treated as centroid of each cluster, next every document is assigned to a nearest cluster based on some similarity measure, then centroids are recomputed. These steps are repeated until there is no change in the centroids for any complete iteration. A contrast and evaluation of these two major categories of document clustering algorithms can be found in [8], which also suggests that Bisecting k-mean is a modified version of k-mean that outperforms AHC in terms of accuracy and is computationally efficient from the quadratic time requirements of AHC. Traditionally, all these document clustering algorithms are based on vector model for computation. A detail on traditional document clustering can be found in [9].

More recently, a new model for document representation has been introduced, based on suffix tree, which is called Suffix Tree Document model. This model is utilized to cluster web documents in [10]. A phrase is an ordered sequence of words, which captures more semantic of text as compared to a single word. Hence, the clustering results produced by phrase based similarity measure are of high quality when compared to the semantic interpretation of the corpus. The more recent work on the phrase based approach is in [11]. A similar work that also utilizes suffix tree model is from [12]. This algorithm starts by computing frequent two-word sets based on user-specified minimum support. Next, all words not in any of the frequent two-word sets are removed from the documents, resulting in compact representation of documents. The compact documents are added one-by-one into a generalized suffix tree data structure. The algorithm then traverses the generalized suffix tree in a depth-first fashion. Every node labeled by a substring of the compact document set (alternatively called a frequent word sequence of the original document set) containing at least two words and supported by at least two documents becomes a cluster candidate. From the set of candidates, it selects the cluster with the longest sequence and merges all clusters with k-mismatched sequences into it. This process is repeated until there are no cluster candidates left. Thus, it produces the clusters. This paper is in fact a comparative study of the algorithms in [11] and [12].

## III. EXPERIMENTAL SET UP

In this section, we evaluate the performance of the two algorithms [11] and [12]. We implemented the algorithms on C# 3.5 and executed the experiments on a Windows 7 based standard PC.

### A. Datasets

Our datasets consist of subsets collected from mix from the standard dataset used for comparing algorithms in document clustering. The OHSUMED collection consists of over 348,566 references from the MEDLINE database, which is a database of medical literature maintained by the National Library of Medicine (NLM). Most of the references have abstracts and all have associated MeSH (Medical Subject Headings) indexing terms, with some of the MeSH terms marked as primary. We generate two datasets from the OHSUMED collection. For the first dataset, we choose the categories used MSH1262, MSH1473, MSH1486, MSH1713, MSH2025 and MSH2030 as identified by theTREC-9 MeSH topics file named"query.mesh.1-4904". For each category, we collect 150 random documents which have the category in their primary MeSH terms, but do not have any of the other chosen categories in their primary MeSH terms. The final dataset contains 1050 documents. We create the second dataset in a similar fashion, but with only 3 categoriesMSH1473, MSH1486, MSH1713. For this dataset, we collect only a 100 documents from each category. The final dataset consists of 300 documents.

The RCV1 (Reuters Corpus Volume 1) contains more than 800,000 manually categorized newswire stories made available by Reuters, Ltd for research. The stories have been manually categorized into at least one topic based on topics from a hierarchy of topic codes. We generate four datasets from the RCV1 collection. For the first dataset, we use the categories C12, C21, C32, E21, E31 and M12. For each category, we collect 60 documents that have the category as their first category and do not have any of the chosen categories among the top two categories. The dataset consists of 360 documents. For the second dataset, we use categories C12, C21 and C32. We collect 100 documents from each category in the same manner as above, forming a dataset of 300 documents. We create the third and fourth datasets in the aforementioned manner using categories C151, C174, C331, E212, E411 and M131 for the third dataset, and C151, E411 and M131 for the fourth. We select 60 documents from each category for the third dataset forming a dataset of 360 documents and select 100 documents from each category for the fourth dataset forming a dataset of 300 documents. Since we have chosen more specific categories for this dataset, we can expect it to be easier to cluster.

All six of the datasets we used are preprocessed before use. Stop words are removed and each word is stemmed using Porter's Suffix Stripping algorithm.

*B. Algorithms*

Neither the source code nor binaries of either algorithm were available; we implemented them ourselves following the description in [11] and [12]. In CFWS, we use the a priori algorithm for finding frequent 2-itemsets and compute the set of frequent words (WS). Then all words not in WS are removed from the documents and the resulting compact documents are added into a generalized suffix tree. Next, we obtain the cluster candidates from the tree based on the k-mismatch concept. We use a simple dynamic programming algorithm for computing all k-mismatched sequences. Hierarchical clustering is then performed on the cluster candidates to obtain the final clustering result. For [11], we first add all documents to a generalized suffix tree, and obtain the feature vector with tf-idf weighting for each document. We only consider the phrases present in at least two documents for constructing the feature vectors since [11] claims that these are the nodes that predominantly determine the result of clustering, and including other nodes only has a slight effect on the result. The documents are then hierarchically clustered using the cosine similarity measure for measuring document similarity and the UPGMA scheme for measuring cluster similarity. The given two diagrams show the steps involved in the two-implemented algorithms.

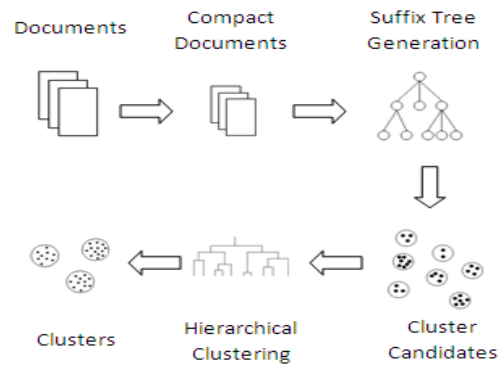

Figure 1 Steps involved in CFWS

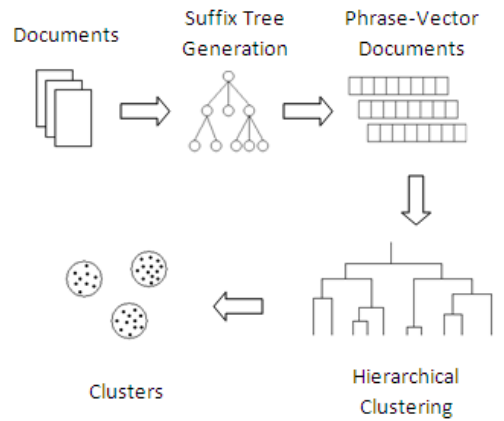

Figure 2: Steps involved in NSTC

*C. Measure*

For measuring the effectiveness of the clustering results, we compare the two algorithms' results on f-measure. Note that we have not used entropy or purity because CFWS allows one document to appear in more than one cluster, and therefore the measure would not be taken for the quality of the result accurately.

The f-measure uses a combination of precision and recall values of clusters. We let $n_i$ designate the number of documents in class $i$, and $c_j$ designate the number of documents in cluster $j$. Moreover, we let $c_{ij}$ designate the number of items of class $i$ present in cluster $j$. Then we can define *prec(i, j)*, the precision of cluster $j$ with respect to class $i$ and *rec(i,j)*, the recall of a cluster $j$ with respect to class $i$ as $prec(i,j) = \frac{c_{ij}}{c_j}$ and $rec(i,j) = \frac{c_{ij}}{n_i}$. The f-measure, *F(i,j)*, of a class $i$ with respect to cluster $j$ is then defined as

$$F(i,j) = \frac{2 * prec(i,j) * rec(i,j)}{prec(i,j) + rec(i,j)}$$

The f-measure for the entire clustering result is defined as

$$F = \sum_i \frac{n_i}{n} \max(F(i,j))$$

For NSTC, we check the best f-score it manages to obtain in the entire hierarchy of clusters it produces. For CFWS, we run the algorithm 3 times; once for each minimum support values of 5%, 6% and 7%. We observe the output hierarchies of clusters and choose the best f-score the algorithm manages to obtain.

*Purity*

Purity can be defined as the maximal precision value for each class j, We compute the purity for a cluster j as $purity(j) = \frac{1}{c_j} max(c_{ij})$. We then define the purity of the entire clustering result as:

$$Purity = \sum_j \frac{c_j}{N} purity(j)$$

Where $N = \sum_j c_j$, i.e. the sum of the cardinalities of each cluster, Note that we use this quantity rather than the size of the document collection for computing the purity.

**Entropy**

Entropy measure how homogenous each cluster j is. It can be calculated by the following formula:

$$Ei = -\sum_{j \in L} presision(i,j) * \log(precision(i,j))$$

The total entropy for a set of cluster is calculated as the sum of entropies for each cluster weighted by the size of each cluster:

$$Entropy_C = \sum_{i \in C} \left(\left(\frac{Ni}{N}\right) * Ei\right)$$

We need to maximize the purity measure and minimize the entropy of clusters in order to accomplish high quality clustering results.

IV. RESULTS

The result of this experiment shows that the F-score obtained from the test data sets clearly exhibits the superiority of algorithm [11] over algorithm [12], on variety of situations. The results reported by respective authors are otherwise of this conclusion. The computational demands of CFWS [12] are much higher than NSTC [11]. The following table plots the F-score of the two algorithms against the test data sets.

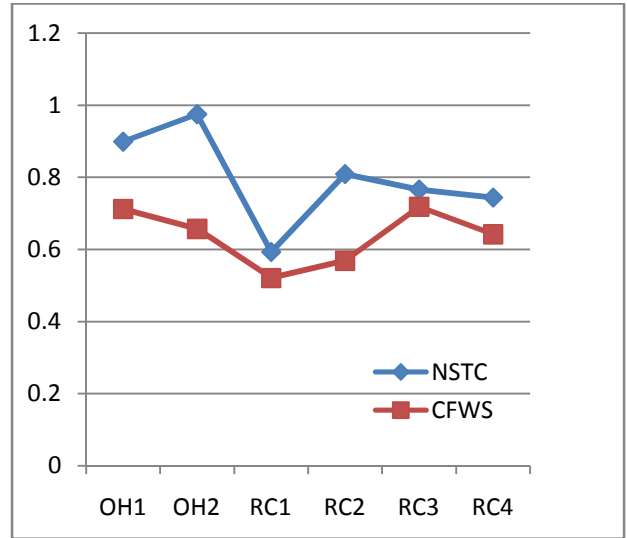

Table 1: F-Score for the 6 datasets

Similarly the purity and entropy of the two algorithms, on the 6 datasets can be seen from the tables given below:

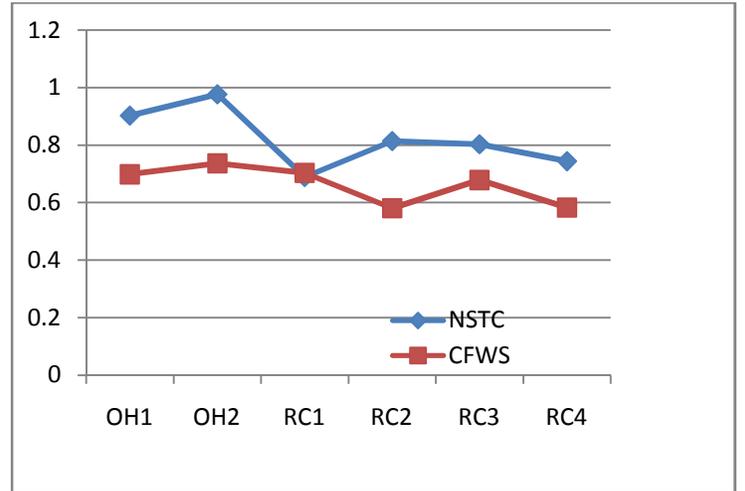

Table 2: F-Score for the 6 datasets

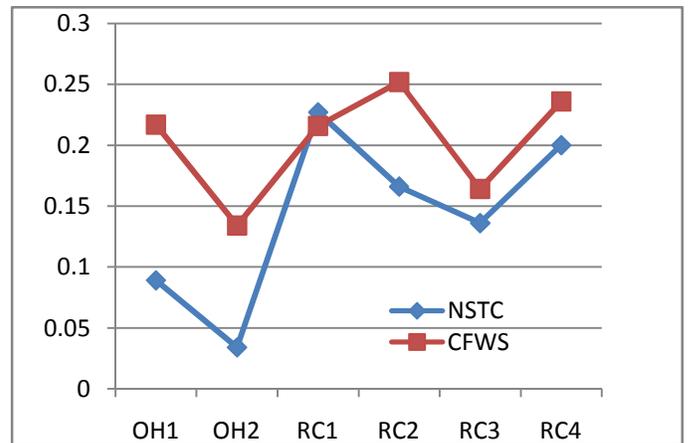

Table 3: Entropy for the 6 datasets

## V. CONCLUSION

It can be clearly concluded from the results obtained that Efficient Phrase based clustering algorithm [NTSC] is superior.